\RequirePackage{lineno}
\documentclass[aps,prl,showpacs,twocolumn,superscriptaddress]{revtex4-1}
\usepackage{amssymb}
\usepackage[final]{graphicx}
\usepackage{amsmath}
\usepackage{epsfig}
\usepackage{color}
\usepackage[figuresright]{rotating}

\usepackage{ulem}
\usepackage{bm}

\begin{document}

\title{A New Clock Transition with the Highest Sensitivity to $\alpha$ Variation and Simultaneous Magic Trapping Conditions with Other Clock Transitions in Yb}

\author{Zhi-Ming Tang}
\email{zmtang18@fudan.edu.cn}
 \affiliation{Shanghai EBIT Laboratory, Key Laboratory of Nuclear Physics and Ion-Beam Application (MOE), Institute of Modern Physics, Fudan University, Shanghai 200433, China}
\author{Yan-mei Yu}
\email{ymyu@aphy.iphy.ac.cn}
 \affiliation{Beijing National Laboratory for Condensed Matter Physics, Institute of Physics, Chinese Academy of Sciences, Beijing 100190, China}
\author{B. K. Sahoo}
 \affiliation{Atomic, Molecular and Optical Physics Division, Physical Research Laboratory, Navrangpura, Ahmedabad 380009, India}
\author{Chen-Zhong Dong}
 \affiliation{Key Laboratory of Atomic and Molecular Physics $\&$ Functional Materials of Gansu Province, College of Physics and Electronic Engineering, Northwest Normal University, Lanzhou 730070, China}
\author{Yang Yang}
\email{yangyang@fudan.edu.cn}
 \affiliation{Shanghai EBIT Laboratory, Key Laboratory of Nuclear Physics and Ion-Beam Application (MOE), Institute of Modern Physics, Fudan University, Shanghai 200433, China}
\author{Yaming Zou}
 \affiliation{Shanghai EBIT Laboratory, Key Laboratory of Nuclear Physics and Ion-Beam Application (MOE), Institute of Modern Physics, Fudan University, Shanghai 200433, China}

\date{August 19, 2022}

\begin{abstract}
Optical lattice clocks are the prospective devices that can probe many subtle physics including temporal variation of the fine structure constant
($\alpha_e$). These studies necessitate high-precision measurements of atomic clock frequency ratios to unprecedented accuracy. In contrast to the
earlier claimed highest sensitive coefficient ($K$) clock transition to $\alpha_e$ in Yb [Phys. Rev. Lett. \textbf{120}, 173001 (2018)], we found the
$4f^{14}6s6p\,(^3P_2) - 4f^{13}5d6s^{2}\,(^3P_2^*)$ transition of this atom can serve as the clock transition with the largest $K$ value (about
$-$25(2)). Moreover, we demonstrate a scheme to attain simultaneous magic trapping conditions for this clock transition with the other two
proposed clock transitions $4f^{14}6s^2\,(^1S_0) - 4f^{14}6s6p\,(^3P_2)$ and $4f^{14}6s^2\,(^1S_0) - 4f^{13}5d6s^{2}\,(^3P_2^*)$; also exhibiting
large $K$ values. This magic condition can be realized by subjecting Yb atoms to a bias magnetic field at a particular polarization angle along the quantization axis in an experimental set up. Upon realization, it will serve as the most potential optical lattice clock to probe $\alpha_e$ variation.
\end{abstract}

\maketitle

There has been a tremendous progress in building up ultra-high precision atomic clocks using optical lattices in the last decade reaching $10^{-18}$ uncertainty level~\cite{Derevianko-RMP-2011, Ludlow-RMP-2015, Campbell-science-2017, Schioppo-nphoton-2017, Oelker-nphoton-2019, Boulder-nature-2021}. These clocks are capable of offering better statistics owing to their large signal-to-noise ratio, hence considered as the future primary frequency standard \cite{Riehle-metrologia-2018}. They operate at magic wavelengths ($\lambda_{\rm magic}$), where the differential Stark shift of a given transition is effectively nullified when the lattice lasers are applied to the atomic systems \cite{Ido-PRL-2003, Ye-science-2008, Brown-PRL-2017}. These clocks are not only meant for time keeping devices, they can also be immensely useful for studying a number of fundamental physics of general interest such as probing variation of fundamental physical constants, relativistic geodesy, gravitational-wave detection, search for dark matter and beyond the Standard Model particle physics to name a few
\cite{Safronova-RMP-2018, McGrew-nature-2018, Grotti-nphys-2018, Kolkowitz-PRD-2016, Derevianko-nphys-2014, Wcislo-sciadv-2018, Kennedy-PRL-2020, Kouvaris-PRD-2021}. In order to suppress statistical noise that limits the stability of the clock frequency measurement, synchronous frequency comparison technique based on a large ensemble of atoms are advantageous
~\cite{Takamoto-nphoton-2011, Zheng-nature-2022, Bothwell-nature-2022}, which can facilitate measurement of the ratio of frequency beyond the Dick
limit \cite{Dick-1987}. However, simultaneously comparing ratios of more than two clock frequencies in the same optical lattice can make these
studies more reliable. This, however, poses many practical challenges for which reducing uncertainties due to light shifts to the intended level
may not be feasible. This, therefore, demands to use a single atomic species and trapping them simultaneously at a common magic wavelength.
Obviously, it is not an easy task to achieve such a goal in an atomic clock candidate straightforwardly but can be feasible by slightly tweaking the experimental set up. For this purpose, we intend to make use of the polarization angle along the quantization axis in the clock frequency measurement by applying a bias magnetic field.

\begin{figure}[t]
\centering
\includegraphics[width=8.6cm]{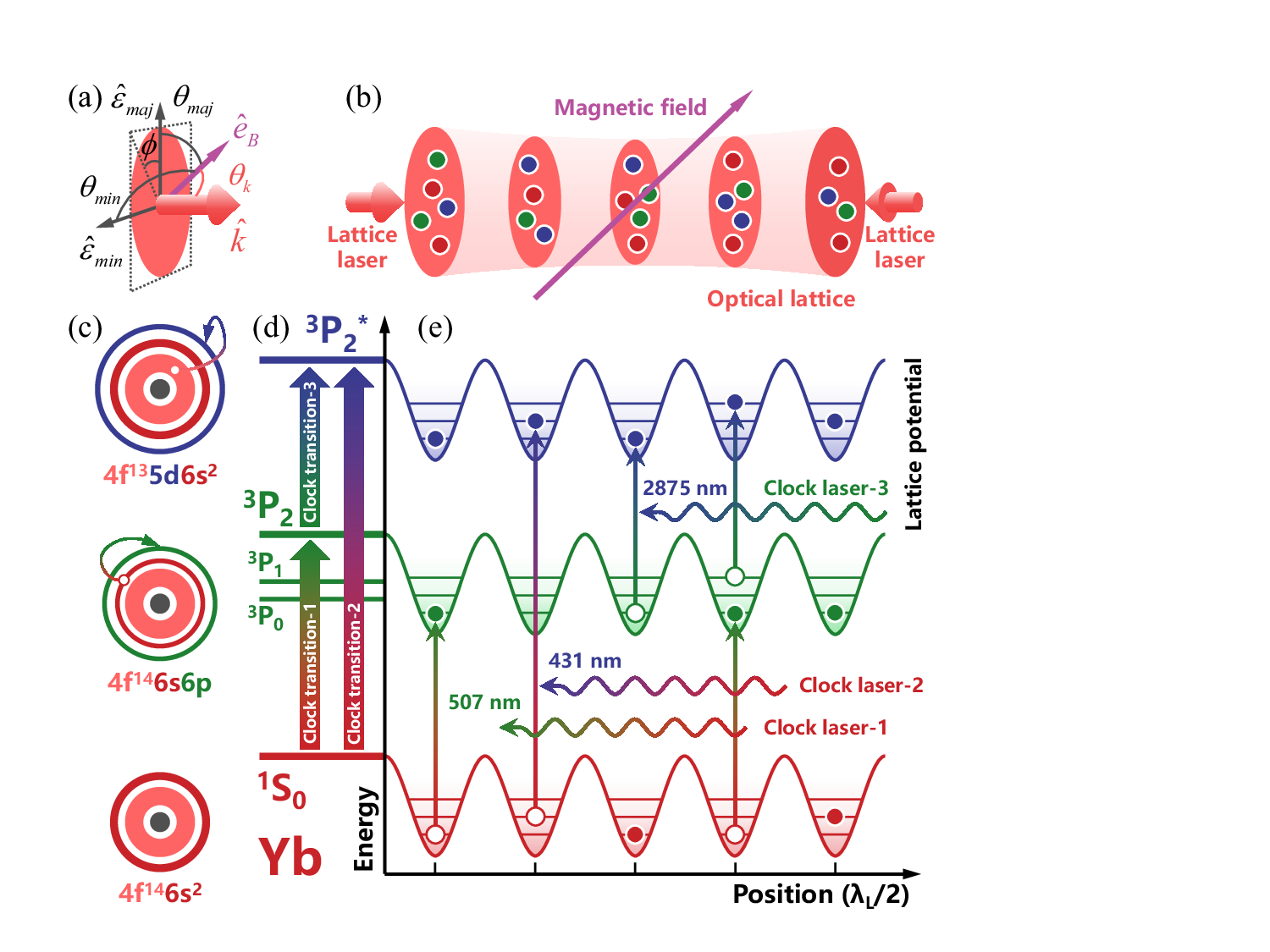}
\caption{Schematic of triple lattice clocks using Yb atoms demonstrating: (a) geometrical configuration of the applied laser with $\bm{\hat{e}_{B}}$ is the quantization axis, $\bm{\hat{k}}$ is the wave vector,  $\phi$ is the polarization angle, $\mathcal{A}=\textmd{sin}2\phi$ denotes degree of polarization and $\bm{\hat{\varepsilon}_{maj}}$ $(\bm{\hat{\varepsilon}_{min}})$ is the major (minor) component of the polarization vector $\bm{\hat{\varepsilon}}$; (b) 1D optical lattice constructed by a standing wave of laser that tuned to a triply magic trapping condition for three states leading to three clock transitions; (c) configurations of the $^1S_0$, $^3P_2$ and $^3P_2^*$ clock states; (d) investigated $^1S_0 - ^3P_2$, $^1S_0 - ^3P_2^*$ and $^3P_2 - ^3P_2^*$ clock transitions; and (e) simultaneous interrogation procedures for all three clock transitions. \label{fig:triclock}}
\end{figure}

The 578 nm line of the $4f^{14}6s^2~(^1S_0)-4f^{14}6s6p~(^3P_0)$ transition in neutral ytterbium (Yb) is currently serving as one of the most
accurate frequency standards \cite{Schioppo-nphoton-2017, Boulder-nature-2021, Brown-PRL-2017}. This transition has a widely-used
$\lambda_{\rm magic}$ at 759 nm \cite{Barber-PRL-2006, Barber-PRL-2008} and several other $\lambda_{\rm magic}$ values in the shorter wavelength
range \cite{Dzuba-JPB-2010}. Recently, it was demonstrated that the transitions $4f^{14}6s6p\,(^3P_0) - 4f^{13}5d6s^{2}\,(^3P_2^*)$,
$4f^{14}6s^2\,(^1S_0) - 4f^{14}6s6p\,(^3P_2)$ and $4f^{14}6s^2\,(^1S_0) - 4f^{13}5d6s^{2}\,(^3P_2^*)$ can be served as additional clock transitions in Yb \cite{Safronova-PRL-2018,Dzuba-PRA-2018} having the highest sensitivity coefficient ($K$) to variation of the fine structure constant ($\alpha_e$).
In this Letter, we demonstrate that the $4f^{14}6s6p\,(^3P_2) - 4f^{13}5d6s^{2}\,(^3P_2^*)$ transition in Yb can suffice as a new clock transition
and has larger $K$ value than the previously claimed highest $K$ value for the $^3P_0 - ^3P_2^*$ clock transition \cite{Safronova-PRL-2018}.
This is further strengthened by finding a triply magic trapping condition to trap and carry out simultaneous clock frequency measurements
for the $^1S_0 - ^3P_2$, $^1S_0 - ^3P_2^*$, and $^3P_2 - ^3P_2^*$ clock transitions on common optical lattice to suppress the Stark shifts
(see Fig.~\ref{fig:triclock}) drastically.

For carrying out the analysis, we have calculated the reduced matrix elements (RMEs) of the electric dipole ($E1$), electric quadrupole ($E2$),
magnetic dipole ($M1$), and magnetic quadrupole ($M2$) operators in Yb using the multiconfiguration Dirac-Hartree-Fock (MCDHF) method
\cite{Grant-springer-2007, Fischer-JPB-2016, Fischer-CPC-2019}. We have considered electron correlation effects due to both the Coulomb and Breit
interactions \cite{Grant-JPB-1976}. Corrections from quantum electrodynamics (QED) are included from the lowest-order self-energy (SE) and the
vacuum polarization (VP) interactions. The SE QED corrections are evaluated based on an expression derived using screened hydrogenic
contributions~\cite{Lowe-RPC-2013}. The VP QED corrections are estimated using an analytical expression derived for the lowest-order term by
Fullerton and Rinker for the Uehling model potential and for the next leading-order term as given in Ref. \cite{Fullerton-PRA-1976}. Using the
RMEs of the $E2$, $M1$, and $M2$ operators, we have estimated lifetimes of the $^3P_2$ and $^3P_2^*$ metastable states of Yb in order to demonstrate
that the $^3P_2 - ^3P_2^*$ transition is apt to be used for the atomic clock along with other clock transitions. As shown in Fig.~\ref{fig:triclock},
the states associated with this clock transition have open-4f shell configurations. Thus, it is extremely challenging to evaluate atomic wave
functions of these states accurately to which we have considered a very large set of configuration state functions (CSFs) in our MCDHF method to
obtain the results reliably. We find that the lifetime of the $^3P_2$ state is about 16 s against the earlier reported values as
15 s~\cite{Dzuba-PRA-2018} and 14.5 s~\cite{Migdalek-JPB-1991}. Our estimated lifetime of the $^3P_2^*$ state comes out to be 190 s, which agrees
better with the value 200 s reported in Ref. \cite{Dzuba-PRA-2018} but deviates from other {\it ab initio} result 1300 s and semi-empirical value
55 s \cite{Safronova-PRL-2018}. Our investigation reveals that both the $^3P_2$ and $^3P_2^*$ states decay to the $^3P_1$ state predominately
by the $M1$ channel with the branching ratios as 0.93 and 0.98, respectively. Similarly, our calculation offers
lifetime of the $^3P_1$ state as 875 ns, which is consistent with the most precise experimental value 865(6) ns~\cite{Beloy-PRAR-2012} and previous
measurement 850(50) ns~\cite{Bowers-PRA-1996}, while differs considerably from the another theoretical prediction 500 ns reported in Ref.
\cite{Dzuba-PRA-2018}. This justifies about reliability of our calculations using large CSFs in the MCDHF method. Owing to short lifetime, this
state will help in repumping the atoms back to the ground state, and the $^1S_0-^3P_1$ transition can be used for narrow-line cooling to
achieve $\mu$K temperature \cite{Kuwamoto-PRA-1999, Saskin-PRL-2019}. It can be noted that the metastable state $^3P_0$, that is part of the
traditional clock state, will behave as a dark state as it will not play any active role in our scheme. This is because any decay to this state
from  other metastable states involved with the clock transitions are highly forbidden, and any possible leakage to this state is further prohibited
strongly to the ground $^1S_0$ state in the bosonic Yb isotopes. During the repumping process, the fraction of atoms leaked to the $^3P_0$ metastable
state can be calibrated \cite{Khramov-PRL-2014}. Otherwise, a 649 nm laser pulse can be used to pump atoms from this state back to the ground state
via the $4f^{14}6s7s\,(^3S_1)$ state~\cite{Yamaguchi-PRL-2008}.

\begin{table}[b]
\caption{Important properties such as transition channels ($O$), wavelengths ($\lambda$), natural linewidths ($\Gamma$), quality factors ($Q$) and fine structure varying sensitive coefficient $K$ of possible optical clock transitions in Yb. Transitions shown in bold are investigated in the present work among which the $ ^3P_2 - ^3P_2^*$ is proposed in this work. Subscripts $hf$ and $at$ denote hyperfine induced and atomic transitions, respectively.} \label{tab1}
{\setlength{\tabcolsep}{2.8pt}
\begin{tabular}{l cccrr}
\hline\hline
Transition	&	$O$ & $\lambda(\textrm{nm})$	&	$\Gamma(\textrm{mHz})$	&	$Q(10^{16})$	&	$K$~	\\ \hline
$^1S_0 - ^3P_0$   	&	E1$_{hf}$ & 578	&	7.0$^a$ 	&	7.4(6)$^a$	&	0.31(4)	 \\
$^3P_0 - ^3P_2^*$   	&	E2$_{at}$ & 1695	&	0.84	&	21(2)	&	$-$14(1)	 \\
$\bm{^1S_0 - ^3P_2}$   	&	M2$_{at}$ &	507	&	9.7	&	6.1(6)	&	0.55(4)	 \\
$\bm{^1S_0 - ^3P_2^*}$   	&	M2$_{at}$  &	431	&	0.84	&	83(8)	&	$-$3.2(2)	 \\
$\bm{^3P_2 - ^3P_2^*}$   	&	M1$_{at}+$E2$_{at}$ &	2875	&	9.8	&	1.1(1)	&	$-$25(2)	\\
\hline\hline
\end{tabular}}
\leftline{$^a$For $^{171}$Yb~\cite{Xu-PRL-2014}.}
\end{table}

\begin{table}[t]
\caption{$\alpha^S$, $\alpha^V$, and $\alpha^T$ values (in a.u.) at $\omega=0$ and $\omega=1056$ nm of the considered clock states
$4f^{14}\,6s^2\,(^1S_0$), $4f^{14}\,6s6p\,(^3P_2$), and $4f^{13}\,6s^2 5d\,(^3P_2^*)$ states of Yb. These dynamic values offers simultaneous
magic trapping conditions for the triple clock transitions.}
\label{tab2}
{\setlength{\tabcolsep}{4.0pt}
\begin{tabular}{l lccl}
\hline\hline
$\alpha$ 	&	$^1S_0$	& $^3P_2$	& $^3P_2^*$	&	Reference \\	
\hline
$\alpha^S(0)$	&	139(3)	& 	409(34)	& 122(6) &	This work		\\
     & 150 &  418   & 124   &  Ref.~\cite{Dzuba-PRA-2018}  \\
      & 139.3(5.0) & & & Ref.~\cite{Beloy} \\
$\alpha^T(0)$ &   &  $-74(6)$ & $-0.04(1.00)$ & This work \\
     &  &  $-70$  &  $-6$  &  Ref.~\cite{Dzuba-PRA-2018}  \\
$\alpha^S(1056)$  & 160(3)  & 133(13) & 149(17) & This work \\
$\alpha^V(1056)$  &  & $-510(57)$ & $-210(31)$  & This work \\
$\alpha^T(1056)$  & & $-98(2)$ & $-39(5)$ & This work \\
\hline\hline
\end{tabular}}
\end{table}

The important properties of the relevant clock transitions in Yb are summarized in Table \ref{tab1}. The $^3P_0 - {^3}P_2^*$ transition was
demonstrated by Safronova $et~al.$ \cite{Safronova-PRL-2018} for development of a dual fermionic clock combining with the traditional
$^1S_0 - ^3P_0$ clock transition. The linewidth of this clock transition, however, can be increased by the hyperfine quenching. In contrast, we
suggest to use the $^3P_0 - ^3P_2^*$ transition in the bosonic Yb isotope for the atomic clock which can have a narrow linewidth due to the
forbidden $E2$ channel. The choice of bosonic isotope can offer enhanced quality ($Q$) factor due to the fact that it will be free from the hyperfine
quenching (nonetheless this transition in a fermionic Yb isotope can still offer a reasonably large $Q$ factor). As stated earlier, simultaneous
magic conditions for the $^1S_0 - ^3P_0$ and $^3P_0 - ^3P_2^*$ clock transitions do not seem to be feasible. On the other hand, the $^1S_0 - ^3P_2$
transition in bosonic Yb has similar natural linewidth and $Q$ factor with the traditional $^1S_0 - ^3P_0$ clock transition in $^{171}\textrm{Yb}$.
In addition, it has a larger $K$ value to probe $\alpha_e$ variation. The $^1S_0 - ^3P_2^*$ transition has even larger $Q$ and $K$ factors. Compared
to these earlier proposed clock transitions, we find that the $^3P_2 - ^3P_2^*$ transition has the largest $K$ factor (about 80 times larger than
the $^1S_0 - ^3P_0$ transition). Also, owing to its 9.8 mHz narrow linewidth, the 2875 nm infrared wavelength possesses $Q$ factor about $1.1 \times
10^{16}$. Neither this value of $Q$ nor wavelength of the laser that will be applied to interrogate the clock transition would limit the stability
of our proposed atomic clock. These analyses suggest that the $^1S_0 - ^3P_2$, $^1S_0 - ^3P_2^*$ and $^3P_2 - ^3P_2^*$ transitions can be as
competitive as the $^1S_0 - ^3P_0$ transition for clock frequency measurements in Yb. Our MCDHF calculation gives $K$ factor for the
$^3P_2 - ^3P_2^*$ transition as $-$25(2). Using empirical relations for the $K$ factors of the $^1S_0 - ^3P_2$ and $^1S_0 - ^3P_2^*$ transitions
from Refs. \cite{Dzuba-PRA-2018, Safronova-PRL-2018}, we find $K$ value for the $^3P_2 - ^3P_2^*$ transition as $-$28.6(2.0) agreeing well with our
result. We also suggest that the $^1S_0 - ^3P_2$ and $^1S_0 - ^3P_2^*$ clock transitions can work as synthetic frequency standard
\cite{Yudin-PRL-2011, Yudin-PRA-2016} as $\alpha_e$ varying coefficients in these transitions have opposite signs. Thus, the $^3P_2 - ^3P_2^*$
transition bestows to offer us a new direction towards development of next-generation high-precision frequency standards.

Our next utmost requirement would be to assess simultaneous magic $\lambda_{\rm magic}$ values for the aforementioned three clock transitions to minimize uncertainties in the clock frequency measurements, which requires accurate determination of dynamic $E1$ polarizabilities ($\alpha (\omega)$) of the considered states of Yb for a wide range of angular frequency $\omega$.

Expression for $\alpha (\omega)$ is given by \cite{Manakov-physrep-1986, Kien-EPJD-2013}
\begin{eqnarray} \label{eq:alphaTotal}
\alpha (\omega) = \alpha^{S}(\omega)+\mathcal{C}_{JM}^{V}(\mathcal{A},\theta_{k})\alpha^{V}(\omega)+ \mathcal{C}_{JM}^{T}(\theta_{p})
\alpha^{T}(\omega),
\end{eqnarray}
where $\alpha^{S}$, $\alpha^{V}$ and $\alpha^{T}$ are known as the scalar, vector and tensor components, respectively, and $\mathcal{C}_{JM}^{V}(\mathcal{A},\theta_{k})$ and $\mathcal{C}_{JM}^{T}(\theta_{p})$ are two polarization-angle-dependent coefficients and given by
\begin{eqnarray} \label{eq:CV}
&& \mathcal{C}_{JM}^{V}(\mathcal{A},\theta_{k}) = \mathcal{A}\textmd{cos}\theta_{k}\frac{M}{2J}, \\
\text{and} && \nonumber \\
&& \mathcal{C}_{JM}^{T}(\theta_{p}) = \frac{3\textmd{cos}^2\theta_{p}-1}{2} \frac{3M^2-J(J+1)}{J(2J-1)}.
\label{eq:CT}
\end{eqnarray}
Here $\mathcal{A}$ represents the degree of polarization defined by polarization angle $\phi$ as $\mathcal{A}=\textmd{sin}2\phi$, $\theta_{k}$ is
the angle between the wave vector of the light and the quantization axis chosen as the direction of the magnetic field, $\theta_{p}$ is the angle
between the polarization vector and the quantization axis, and $M$ is the projection of the total angular momentum $J$ at the quantization axis.
Using the calculated $E1$ RMEs and experimental energies \cite{NIST}, the dominant valence electron correlation contributions to each component of
$\alpha (\omega)$  are determined. The other contributions from the core and core-valence electron correlations are estimated using mixed many-body
methods. We have listed their static ($\omega=0$) and dynamic ($\omega=1056$ nm) values along with uncertainties of the $^1S_0$, $^3P_2$ and
$^3P_2^*$ states from our calculations in Table \ref{tab2}. We have also compared the static values with the latest theoretical calculation
\cite{Dzuba-PRA-2018} and another work that constraints the value by analyzing several $E1$ matrix elements using the experimental data \cite{Beloy}.
We find overall agreement among our results with the previous calculation
\cite{Dzuba-PRA-2018} except for the $\alpha^T$ value of the $^3P_2^*$ state ($-$0.04(1.00) a.u. and $-$6 a.u.). After a thorough investigation we
observe that the large difference mainly comes due to different order of contributions arising from the low-lying resonant transitions. We find that
the first six resonances that have the transition energies between 1300-9506 cm$^{-1}$ give almost all contributions to the determination
of $E1$ RMEs. Among these six states, the most dominant contribution to the $\alpha^T$ value comes from the $^3P_2^* - 4f^{13}\,6s^2\,6p~J=3$
resonance line, amounting to $-$1.25 a.u.. Consideration of very large number of CSFs in our MCDHF calculations improves the accuracy of the
transition energies. The difference between our calculated values and the experimental results \cite{NIST} is reduced to be less than 200 cm$^{-1}$
compared to $>$1000 cm$^{-1}$ in Ref. \cite{Dzuba-PRA-2018}. Using the calculated static $\alpha^{S}$ values of the $^1S_0$, $^3P_2$ and $^3P_2^*$ states of Yb from this work, we estimate the black-body
radiation (BBR) shifts at the room temperature of the $^1S_0-^3P_2$, $^1S_0-^3P_2^*$ and $^3P_2-^3P_2^*$ clock transitions as $3.9(4)\times10^{-15}$,
$2.1(3)\times10^{-16}$ and $2.4(2)\times10^{-14}$, respectively. When the instrument temperature is well-controlled, for instance, fluctuation in
the temperature can be restricted within mK using techniques like in-vacuum thermal shields~\cite{Beloy-PRL-2014, McGrew-nature-2018, Heo-arxiv-2022},
uncertainties in the BBR shifts for the $^1S_0-^3P_2$, $^1S_0-^3P_2^*$ and $^3P_2-^3P_2^*$ clock transitions can be suppressed below $10^{-18}$
levels easily. At this stage, we would also like to mention that quadratic Zeeman shift can be small \cite{Dzuba-PRA-2018} whereas quadrupole
shift can be nullified by measuring clock frequencies in all the azimuthal levels in the $^3P_2-^3P_2^*$ transition.

\begin{figure}[t]
\centering
\includegraphics[width=4.27cm,clip,trim={0cm 0cm -0.03cm 0cm}]{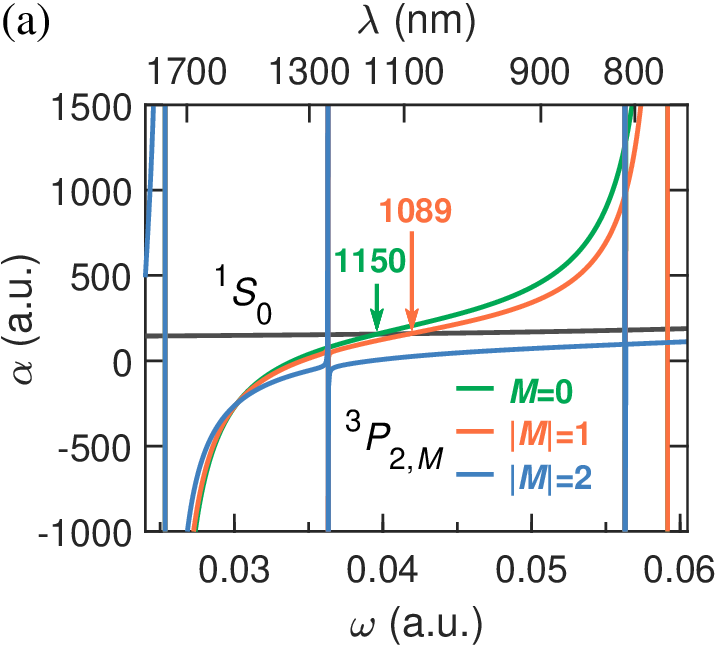}  \hspace{-5pt}
\includegraphics[width=4.10cm,clip,trim={0cm 0cm -0.03cm 0cm}]{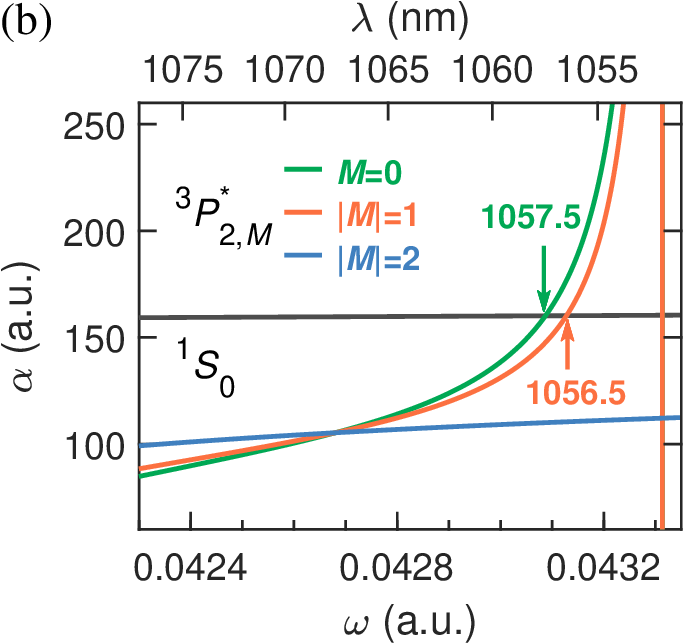}  \\  \vspace{4pt} \hspace{-1pt}
\includegraphics[width=4.25cm,clip,trim={0cm 0cm -0.03cm 0cm}]{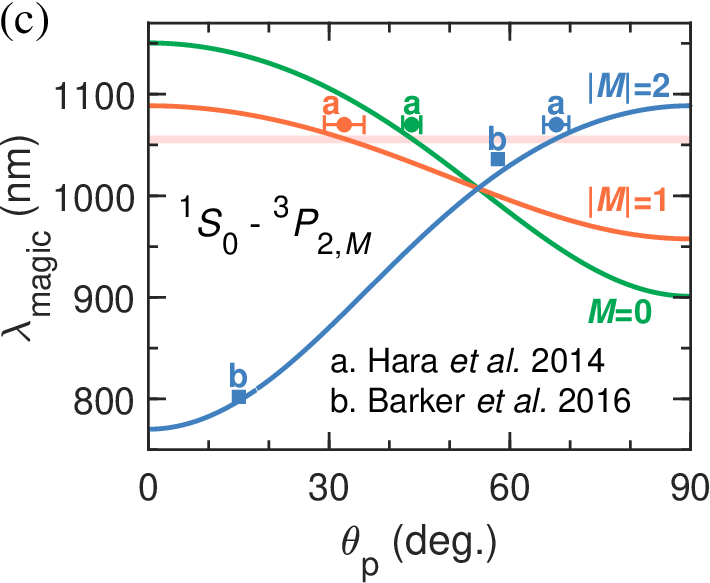}  \hspace{-5.5pt}
\includegraphics[width=4.25cm,clip,trim={0cm 0cm -0.03cm 0cm}]{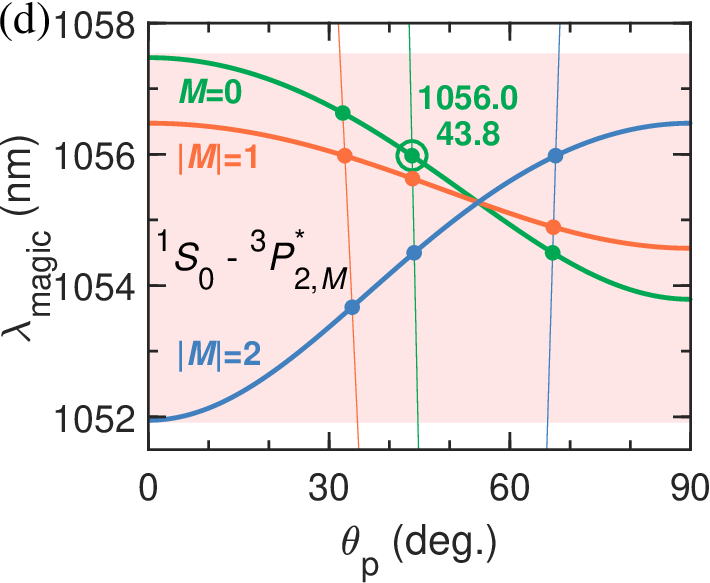}  \\
\setlength{\abovecaptionskip}{8pt}
\caption{(a)-(b) Dynamic polarizabilities $\alpha(\omega)$ of the $4f^{14}6s^2\,(^1S_0)$, $4f^{14}6s6p\,(^3P_2)$ and $4f^{13}5d6s^{2}\,(^3P_2^*)$ clock states of Yb in the linearly ($\pi$) polarized light at $\theta_p=0^{\circ}$.
Vertical arrows denotes the far-off-resonance magic wavelengths.
(c)-(d) $\theta_p$-dependent magic wavelengths in the $\pi$ polarized light for the $^1S_0-^3P_2$ and $^1S_0-^3P_2^*$ clock transitions.
The triply magic trapping conditions are indicated by dots in (d).
The green circle highlights the triply magic wavelength condition for $M=0$ sublevels at $\lambda_{\textrm{3magic}}=1056.0~\textrm{nm}$ and $\theta_p=43.8^{\circ}$.
References in (c): $^a$Experiment~\cite{Hara-JPSJ-2014},$^b$Theory~\cite{Barker-PRA-2016}.
\label{fig:magicpi} }
\end{figure}

\begin{figure}[t]
\centering
\includegraphics[width=6.5cm,clip,trim={0cm 0cm -0.1cm 0cm}]{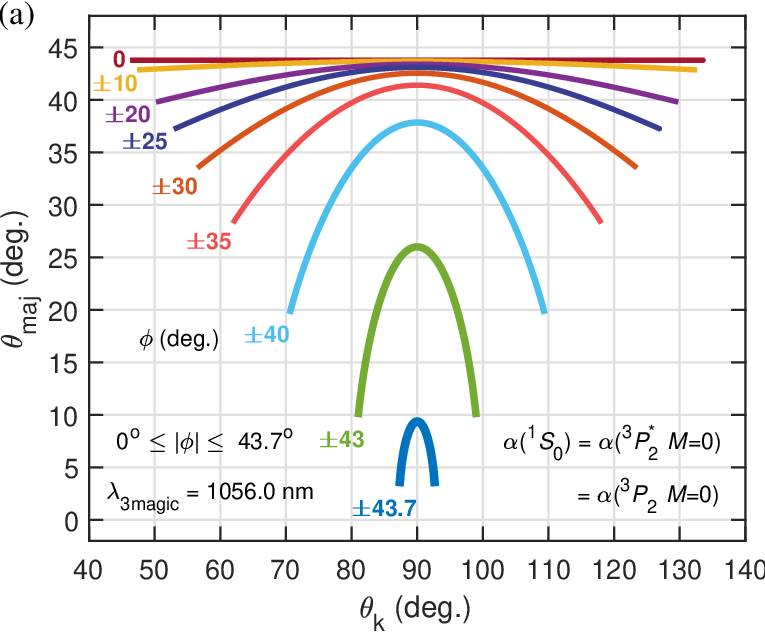} \\  \vspace{6pt} \hspace{4pt}
\includegraphics[width=7.0cm,clip,trim={0cm 0cm 0.3cm 0.3cm}]{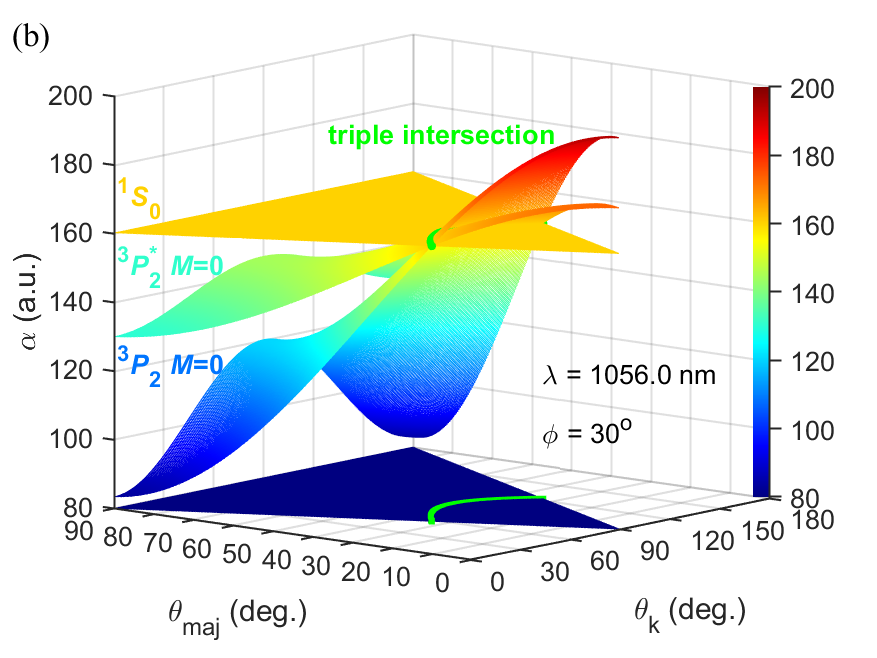}  \\
\setlength{\abovecaptionskip}{8pt}
\caption{(a) Triply magic trapping conditions for the $^1S_0$, $^3P_2~M=0$ and $^3P_2^*~M=0$
clock states of Yb in different elliptically polarized light. The conditions can be found
for $0^{\circ} \leqslant |\phi| \leqslant 43.7^{\circ}$.
In the case of all $M=0$, the common magic wavelength $\lambda_{\textrm{3magic}}=1056.0~\textrm{nm}$
is independent with polarization angles.
(b) $\{\theta_{maj},\theta_k\}$-dependent dynamic polarizabilities of the $^1S_0$, $^3P_2~M=0$ and $^3P_2^*~M=0$ sublevels at $\lambda_{\textrm{3magic}}=1056.0~\textrm{nm}$, by choosing $\phi=30^{\circ}$ as an example.
The surface shape for other polarized light is similar.
The green curved line indicates the triple intersection of three surfaces for polarizabilities of three sublevels,
where the triply magic trapping conditions are found. \label{fig:magicellipt} }
\end{figure}

In the search of $\lambda_{\textrm{magic}}$, we start with the simple case of linearly polarized light and plot the $\alpha(\omega)$ values for the
$^1S_0-^3P_2$ transition (Fig. \ref{fig:magicpi} (a)) and the $^1S_0-^3P_2^*$ transition (Fig. \ref{fig:magicpi} (b)) with $\theta_p=0^{\circ}$ and
$\phi=0^{\circ}$. Fig. \ref{fig:magicpi} (a) shows that the far-off-resonance $\lambda_{\textrm{magic}}$ for the $^1S_0-^3P_2$ transition occurs at
1150 nm for the $M=0$ sublevel and 1089 nm for the $|M|=1$ sublevel. Fig. \ref{fig:magicpi} (b) shows that there exists a $\lambda_{\textrm{magic}}$
for the the $^1S_0-^3P_2^*$ transition at 1057.5 nm for the $M=0$ sublevel and 1056.5 nm for the $|M|=1$ sublevel, respectively, which are close to the above $\lambda_{\textrm{magic}}$. In the case of the linearly polarized light,
$\theta_p$ is a free parameter that can be monitored to adjust the place of the $\lambda_{\textrm{magic}}$ values. In Figs.
\ref{fig:magicpi} (c) and (d), we show variation of $\lambda_{\textrm{magic}}$ as function of $\theta_p$ for the $^1S_0-^3P_2$ and $^1S_0-^3P_2^*$
clock transitions. The dot mark with values `1056.0, 43.8' shown in Fig. \ref{fig:magicpi} (d) indicates a simultaneous magic condition for
the $^1S_0-^3P_2$, $^1S_0-^3P_2^*$ and $^3P_2-^3P_2^*$ clock transitions with $M=0$ sublevels. The reason to use the $M=0$ sublevels of the $^3P_2$
and $^3P_2^*$ states in the clock transition is to nullify the linear Zeeman shift. A similar exercise has been carried out to search for
$\lambda_{\textrm{magic}}$ values for our interested clock transitions in Yb using circularly polarized light. However, we did not find any
simultaneous $\lambda_{\textrm{magic}}$ values for the $M=0$ sublevels of the $^3P_2$ and $^3P_2^*$ states, {\it albeit} they can exist for other
$M$ sublevels.

As seen above we had a limited scope to make use of simultaneous $\lambda_{\textrm{magic}}$ for the three $^1S_0-^3P_2$, $^1S_0-^3P_2^*$ and
$^3P_2-^3P_2^*$ clock transitions in Yb using the linearly and circularly polarized lasers. To overcome this restriction, we further explore the
use of elliptically polarized light to achieve more useful $\lambda_{\textrm{magic}}$ values to carry out the experiments, as shown in
Fig. \ref{fig:magicellipt}. Again, we intend to focus only on the $M=0$ sublevels of the clock transitions in order to suppress linear Zeeman
effects. This choice can also be helpful to avoid drifting and imperfection in achieving polarization angles as well as guiding direction of the
magnetic field. Since the polarization vector is divided into two components as $\bm{\hat{\varepsilon}} = e^{\textmd{i}\sigma} ( \textmd{cos}
\phi \bm{\hat{\varepsilon}_{maj}} + \textmd{i} ~ \textmd{sin}\phi \bm{\hat{\varepsilon}_{min}} )$, $\theta_p$ can be defined as
$\textmd{cos}^2\theta_p =  \sqrt{1-{\mathcal{A}}^2} \textmd{cos}^2\theta_{maj} + \frac{1-\sqrt{1-{\mathcal{A}}^2}}{2} \textmd{sin}^2\theta_k$.
Thus, we can only have the independent parameters as $\phi$, $\theta_k$ and $\theta_{maj}$ to manipulate the $\alpha(\omega)$ values as per our
experimental need. Fig.~\ref{fig:magicellipt} (a) shows that it is possible to attain triply magic trapping condition at 1056.0 nm
for a suitable of combinations of $\theta_k$ and $\theta_{maj}$ by varying them in a range of $0^{\circ} \leqslant \phi \lesssim 43.7^{\circ}$. This clearly demonstrates that experimental desired triply magic wavelength conditions can be easily engineered by
using elliptically polarized light than a $\pi$-polarized light. This is owing to the fact that when $\phi$ is close to $0^{\circ}$,
controlling $\phi$ becomes more challenging for a suitable value of $\theta_{maj}$ to attain $\lambda_{\textrm{magic}}$. In contrast, when
$\phi$ is close to $43.7^{\circ}$, fixing $\theta_{maj}$ becomes extremely sensitive to attain $\lambda_{\textrm{magic}}$. In order to avoid
these uncertainties, we suggest to consider moderate values of $\phi$ for triply magic conditions of the atomic clocks. On the other hand,
it would be easier to control $\theta_{k}$ than $\phi$ and $\theta_{maj}$ in an experiment. To circumvent these problems, we recommend
to use $\phi=30^{\circ}$ and $\theta_{k}=90^{\circ}$, as depicted in Fig.~\ref{fig:magicellipt} (b), as optimum values of $\phi$ and $\theta_{maj}$
to avoid systematic uncertainties due to their fluctuations. This corresponds to  $|\bm{\hat{\varepsilon}_{maj}}| =
2|\bm{\hat{\varepsilon}_{min}}|$, which is feasible to accomplish with the help of Glan-Taylor laser polarizer and high-precision ellipsometry.

In conclusion, we have proposed an optical clock using the $^3P_2 - ^3P_2^*$ transition possessing the highest sensitivity to the temporal variation
of the fine structure constant compared to the previously undertaken optical clock transitions. By analyzing spectroscopic properties, we have
demonstrated that its quality factor and black-body radiation shift are at par with the other clock transitions of the same atom. Furthermore,
we have given magic conditions for simultaneous clock frequency measurements in the $^1S_0 - ^3P_2$, $^1S_0 - ^3P_2^*$ and $^3P_2 - ^3P_2^*$ clock
transitions which will not only be useful for reducing systematic effects, it can also open up the further scope to probe temporal variation of the fine structure constant and other fundamental physics unambiguously.


This work was supported by the National Natural Science Foundation of China
(Grant Nos. 11704076, 11911530229, U1732140, 2021AMF01002, 1916321TS00103201, 11874064, and 11874051),
the National Key Research and Development Program of China (Approved No. 2020YFB1902100, and Grant No. 2021YFA1402104),
and the IDP of CAS (Approved No.YJKYYQ20180013). BKS acknowledges use of the HPC facility Vikram-100 at the Physical
Research Laboratory, Ahmedabad.


\begin{thebibliography}{}

\bibitem{Derevianko-RMP-2011}
A. Derevianko, and H. Katori, Rev. Mod. Phys. \textbf{83}, 331 (2011).

\bibitem{Ludlow-RMP-2015}
A. D. Ludlow, M. M. Boyd, J. Ye, E. Peik, and P. O. Schmidt, Rev. Mod. Phys. \textbf{87}, 637 (2015).

\bibitem{Campbell-science-2017}
S. L. Campbell, R. B. Hutson, G. E. Marti, A. Goban, N. Darkwah Oppong,
R. L. McNally, L. Sonderhouse, J. M. Robinson, W. Zhang, B. J. Bloom, and J. Ye,
Science \textbf{358}, 90 (2017).

\bibitem{Schioppo-nphoton-2017}
M. Schioppo, R. C. Brown, W. F. McGrew, N. Hinkley, R. J. Fasano, K. Beloy, T. H. Yoon, G. Milani,
D. Nicolodi, J. A. Sherman, N. B. Phillips, C. W. Oates, and A. D. Ludlow,
Nat. Photon. \textbf{11}, 48 (2017).

\bibitem{Oelker-nphoton-2019}
E. Oelker, R. B. Hutson, C. J. Kennedy, L. Sonderhouse, T. Bothwell, A. Goban, D. Kedar,
C. Sanner, J. M. Robinson, G. E. Marti, D. G. Matei, T. Legero, M. Giunta, R. Holzwarth,
F. Riehle, U. Sterr, and J. Ye,
Nat. Photon. \textbf{13}, 714 (2019).

\bibitem{Boulder-nature-2021}
Boulder Atomic Clock Optical Network (BACON) Collaboration, Nature \textbf{591}, 564 (2021).

\bibitem{Riehle-metrologia-2018}
F. Riehle, P. Gill, F. Arias, and L. Robertsson, Metrologia \textbf{55}, 188 (2018).

\bibitem{Ido-PRL-2003}
T. Ido, and H. Katori, Phys. Rev. Lett. \textbf{91}, 053001 (2003).

\bibitem{Ye-science-2008}
J. Ye, H. J. Kimble, and H. Katori, Science \textbf{320}, 1734 (2008).

\bibitem{Brown-PRL-2017}
R. C. Brown, N. B. Phillips, K. Beloy, W. F. McGrew, M. Schioppo, R. J. Fasano, G. Milani, X. Zhang, N. Hinkley, H. Leopardi,
T. H. Yoon, D. Nicolodi, T. M. Fortier, and A. D. Ludlow, Phys. Rev. Lett. \textbf{119}, 253001 (2017).

\bibitem{Safronova-RMP-2018}
M. S. Safronova, D. Budker, D. DeMille, D. F. Jackson Kimball, A. Derevianko, and C. W. Clark,
Rev. Mod. Phys. \textbf{90}, 025008 (2018).

\bibitem{McGrew-nature-2018}
W. F. McGrew, X. Zhang, R. J. Fasano, S. A. Sch\"{a}ffer, K. Beloy, D. Nicolodi,
R. C. Brown, N. Hinkley, G. Milani, M. Schioppo, T. H. Yoon, and A. D. Ludlow,
Nature \textbf{564}, 87 (2018).

\bibitem{Grotti-nphys-2018}
J. Grotti, et al. Nat. Phys. \textbf{14}, 437 (2018).

\bibitem{Kolkowitz-PRD-2016}
S. Kolkowitz, I. Pikovski, N. Langellier, M. D. Lukin, R. L. Walsworth, and J. Ye,
Phys. Rev. D \textbf{94}, 124043 (2016).

\bibitem{Derevianko-nphys-2014}
A. Derevianko, and M. Pospelov, Nat. Phys. \textbf{10}, 933 (2014).

\bibitem{Wcislo-sciadv-2018}
P. Wcis{\l}o, P. Ablewski, K. Beloy, S. Bilicki, M. Bober, R. Brown, R. Fasano, R. Ciury{\l}o,
H. Hachisu, T. Ido, J. Lodewyck, A. Ludlow, W. McGrew, P. Morzy\'{n}ski, D. Nicolodi,
M. Schioppo, M. Sekido, R. Le Targat, P. Wolf, X. Zhang, B. Zjawin, and M. Zawada,
Sci. Adv. \textbf{4}, eaau4869 (2018).

\bibitem{Kennedy-PRL-2020}
C. J. Kennedy, E. Oelker, J. M. Robinson, T. Bothwell, D. Kedar, W. R. Milner, G. E. Marti,
A. Derevianko, and J. Ye, Phys. Rev. Lett. \textbf{125}, 201302 (2020).

\bibitem{Kouvaris-PRD-2021}
C. Kouvaris, E. Papantonopoulos, L. Street, and L. C. R. Wijewardhana,
Phys. Rev. D \textbf{104}, 103025 (2021).

\bibitem{Takamoto-nphoton-2011}
M. Takamoto, T. Takano, and H. Katori, Nat. Photon. \textbf{5}, 288 (2011).

\bibitem{Zheng-nature-2022}
X. Zheng, J. Dolde, V. Lochab, B. N. Merriman, H. Li, and S. Kolkowitz, Nature \textbf{602}, 425 (2022).

\bibitem{Bothwell-nature-2022}
T. Bothwell, C. J. Kennedy, A. Aeppli, D. Kedar, J. M. Robinson, E. Oelker, A. Staron,
and J. Ye, Nature \textbf{602}, 420 (2022).

\bibitem{Dick-1987}
J. G. Dick, in \textit{Proceedings of the 19th Precise Time and Time Interval Applications and Planning Meeting},
133-147 (1987).

\bibitem{Barber-PRL-2006}
Z. W. Barber, C. W. Hoyt, C. W. Oates, L. Hollberg, A. V. Taichenachev, and V. I. Yudin,
Phys. Rev. Lett. \textbf{96}, 083002 (2006).

\bibitem{Barber-PRL-2008}
Z. W. Barber, J. E. Stalnaker, N. D. Lemke, N. Poli, C. W. Oates, T. M. Fortier, S. A. Diddams, L. Hollberg,
C. W. Hoyt, A. V. Taichenachev, and V. I. Yudin, Phys. Rev. Lett. \textbf{100}, 103002 (2008).

\bibitem{Dzuba-JPB-2010}
V. A. Dzuba, and A. Derevianko, J. Phys. B: At. Mol. Opt. Phys. \textbf{43}, 074011 (2010).

\bibitem{Safronova-PRL-2018}
M. S. Safronova, S. G. Porsev, C. Sanner, and J. Ye, Phys. Rev. Lett. \textbf{120}, 173001 (2018).

\bibitem{Dzuba-PRA-2018}
V. A. Dzuba, V. V. Flambaum, and S. Schiller, Phys. Rev. A \textbf{98}, 022501 (2018).

\bibitem{Grant-springer-2007}
I. P. Grant, \textit{Relativistic Quantum Theory of Atoms and Molecules: Theory and Computation}, (Springer, New York, 2007).

\bibitem{Fischer-JPB-2016}
C. Froese Fischer, M. Godefroid, T. Brage, P. J$\ddot{o}$nsson, and G. Gaigalas, J. Phys. B \textbf{49}, 182004 (2016).

\bibitem{Fischer-CPC-2019}
C. Froese Fischer, G. Gaigalas, P. J\"{o}nsson, and J. Biero\'{n}, Comput. Phys. Comm. \textbf{237}, 184 (2019).

\bibitem{Grant-JPB-1976}
I. P. Grant, and N. C. Pyper, J. Phys. B: At. Mol. Opt. Phys. \textbf{9}, 761 (1976).

\bibitem{Lowe-RPC-2013}
J. A. Lowe, C. T. Chantler, and I. P. Grant, Radiat. Phys. Chem. \textbf{85}, 118 (2013).

\bibitem{Fullerton-PRA-1976}
L. W. Fullerton, and G. A. Rinker Jr, Phys. Rev. A \textbf{13}, 1283 (1976).

\bibitem{Migdalek-JPB-1991}
J. Migdalek, and W. E. Baylis, J. Phys. B: At. Mol. Opt. Phys. \textbf{24}, 99 (1991).

\bibitem{Beloy-PRAR-2012}
K. Beloy, J. A. Sherman, N. D. Lemke, N. Hinkley, C. W. Oates, and A. D. Ludlow,
Phys. Rev. A \textbf{86}, 051404(R) (2012).

\bibitem{Bowers-PRA-1996}
C. J. Bowers, D. Budker, E. D. Commins, D. DeMille, S. J. Freedman, A.-T. Nguyen, S.-Q. Shang, and M. Zolotorev,
Phys. Rev. A \textbf{53}, 3103 (1996).

\bibitem{Kuwamoto-PRA-1999}
T. Kuwamoto, K. Honda, Y. Takahashi, and T. Yabuzaki, Phys. Rev. A \textbf{60}, 745(R) (1999).

\bibitem{Saskin-PRL-2019}
S. Saskin, J. T. Wilson, B. Grinkemeyer, and J. D. Thompson, Phys. Rev. Lett. \textbf{122}, 143002 (2019).

\bibitem{Khramov-PRL-2014}
A. Khramov, A. Hansen, W. Dowd, R. J. Roy, C. Makrides, A. Petrov, S. Kotochigova, and S. Gupta,
Phys. Rev. Lett. \textbf{112}, 033201 (2014).

\bibitem{Yamaguchi-PRL-2008}
A. Yamaguchi, S. Uetake, D. Hashimoto, J. M. Doyle, and Y. Takahashi, Phys. Rev. Lett. \textbf{101}, 233002 (2008).

\bibitem{Xu-PRL-2014}
C.-Y. Xu, J. Singh, J. C. Zappala, K. G. Bailey, M. R. Dietrich, J. P. Greene, W. Jiang, N. D. Lemke,
Z.-T. Lu, P. Mueller, and T. P. O'Connor, Phys. Rev. Lett. \textbf{113}, 033003 (2014).

\bibitem{Yudin-PRL-2011}
V. I. Yudin, A. V. Taichenachev, M. V. Okhapkin, S. N. Bagayev, Chr. Tamm, E. Peik, N. Huntemann,
T. E. Mehlst\"{a}ubler, and F. Riehle, Phys. Rev. Lett. \textbf{107}, 030801 (2011).

\bibitem{Yudin-PRA-2016}
V. I. Yudin, A. V. Taichenachev, M. Yu. Basalaev, and T. Zanon-Willette, Phys. Rev. A \textbf{94}, 052505 (2016).

\bibitem{Manakov-physrep-1986}
N. L. Manakov, V. D. Ovsiannikov, and L. P. Rapoport, Rhys. Rep. \textbf{141}, 320 (1986).

\bibitem{Kien-EPJD-2013}
F. L. Kien, P. Schneeweiss, and A. Rauschenbeutel, Eur. Phys. J. D  \textbf{67}, 92 (2013).

\bibitem{Beloy}
K. Beloy, Phys. Rev. A {\bf 86}, 022521 (2012).

\bibitem{NIST}
A. Kramida, Yu. Ralchenko, J. Reader, and NIST ASD Team (2021). \textit{NIST Atomic Spectra Database} (ver. 5.9), [Online].
Available: https://physics.nist.gov/asd. National Institute of Standards and Technology, Gaithersburg, MD.

\bibitem{Beloy-PRL-2014}
K. Beloy, N. Hinkley, N. B. Phillips, J. A. Sherman, M. Schioppo, J. Lehman, A. Feldman, L. M. Hanssen, C. W. Oates, and A. D. Ludlow,
Phys. Rev. Lett. \textbf{113}, 260801 (2014).

\bibitem{Heo-arxiv-2022}
M.-S. Heo, H. Kim, D.-H. Yu, W.-K. Lee, and C. Y. Park, arXiv:2207.07322 (2022).

\bibitem{Hara-JPSJ-2014}
H. Hara, H. Konishi, S. Nakajima, Y. Takasu, and Y. Takahashi, J. Phys. Soc. Jpn. \textbf{83}, 014003 (2014).

\bibitem{Barker-PRA-2016}
D. S. Barker, N. C. Pisenti, B. J. Reschovsky, and G. K. Campbell, Phys. Rev. A \textbf{93}, 053417 (2016).


\end{thebibliography}
\end{document}